\documentclass{ptephy_v1}%%%%%% to generate preprint number with ptep logo

\preprintnumber{XXXX-XXXX} %%% %%% Insert preprint number here

\usepackage{amsmath}% for dealing with mathematics,
\usepackage{graphics}% for dealing with figures.
\usepackage{subfigure} %for getting the subfigures e.g., "Figure 1a and 1b" etc.

\begin{document}

\title{A Fast High-Voltage Switching Multiwire Proportional Chamber}

%%%% To generate auto affiliation numbers please use \author{}\affil{} command

\author[1\thanks{Present Address: Institute for Basic Science (IBS), Daejeon, Korea 34141, Korea}*]{H. Natori}
\author[2]{N.~Teshima}
\author[3]{M.~Aoki}
\author[1]{H.~Nishiguchi}
\author[3]{T.~D.~Nguyen}
\author[2]{Y.~Takezaki}
\author[2]{Y.~Furuya}
\author[3\thanks{Present Address: Okayama University, Okayama 700-8530, Japan}]{S.~Ito}
\author[1]{S.~Mihara}
\author[3]{D.~Nagao}
\author[1\thanks{Present Address: Institute of High Energy Physics (IHEP), Beijing, 100049, China}]{Y.~Nakatsugawa}
\author[3]{T.~M.~Nguyen}
\author[2]{Y.~Seiya}
\author[2]{K.~Shimizu}
\author[2]{K.~Yamamoto}

\affil[1]{\normalsize \it High Energy Accelerator Research
Organization (KEK), Ibaraki 305-0801, Japan}
\affil[2]{\normalsize \it Osaka City University, Osaka 558-8585, Japan}
\affil[3]{\normalsize \it Osaka University, Osaka 560-0043, Japan}
\affil[*]{E-mail: natori@ibs.re.kr}

\begin{abstract}
A new experiment, called DeeMe, which is designed to search for $\mu$-$e$ conversions with a sensitivity of $\mathcal{O}(10^{-14})$, is in preparation
at the Japan Proton Accelerator Research Complex (J-PARC).
It utilizes a high-quality pulsed proton beam from
the Rapid Cycling Synchrotron at J-PARC.
The detector for DeeMe must tolerate large pulses of prompt charged particles
whose instantaneous hit rate is as large as 70~GHz/mm$^2$ in a time width of
200~ns
and detect a single electron that arrives with delayed timing on
the order of microseconds.
A special wire chamber has been developed with a new
dynamic gain control technique that reduces space charge effects.
In this paper, we detail the novel detector scheme and operation verification.
\end{abstract}

\subjectindex{C31, H11}

\maketitle

\section{Introduction}
We are developing a new experiment, called DeeMe \cite{DeeMe}, to search for $\mu$-$e$
conversions with a sensitivity of $\mathcal{O}(10^{-14})$,
at the Material and Life Science Experimental Facility (MLF) of the
Japan Proton Accelerator Research Complex (J-PARC).
We utilize a high-quality pulsed proton beam from the Rapid Cycling Synchrotron
(RCS) ring \cite{Acc},
in which two bunches of protons are accelerated to 3~GeV
and transported to the MLF by fast extraction.
The two bunches of protons, separated by 600~ns, each with a time width of 200~ns,
hit a production target in the muon science facility (MUSE) of MLF with a repetition rate of 25~Hz.
Pions produced by the collisions decay to muons,
some of which stop inside the target.
Negative muons are trapped in the atoms of the target material to form muonic
atoms and $\mu$-$e$ conversions may occur.
We capture electrons from the production target with
H-line, one of the secondary beamlines,
and search for $\mu$-$e$ conversions, which are
characterized by electrons with a typical momentum of 105~MeV/$c$ arriving approximately
a microsecond after the prompt particles.

The initial proton pulses produce a large number of charged particles and these
prompt burst pulses pass the DeeMe detector,
where a set of multiwire proportional chambers (MWPCs) serves as the primary detector
component.
The number of charged particles
 produced by a proton beam with a power of
1 MW is estimated, using G4beamline \cite{G4beamline}, to be $7.8 \times 10^7$ and $1.9 \times 10^8$ per pulse
at the position of the MWPC,
covering an effective area of 250~mm~$\times$~200~mm,
for carbon and silicon carbide production targets,
respectively.
G4beamline is a particle tracking simulation program based on Geant4 \cite{Geant4}
and is optimized for simulating beamlines.
From this calculated beam profile,
we estimate the instantaneous hit rate per area at the position of the first chamber
to be at most 70~GHz/mm$^2$ at the beam center.
After the burst pulses, the $\mu$-$e$ conversion electrons arrive at the detector with average delays
 of 2.0~$\mu$s and 0.76~$\mu$s
for the C and SiC targets, respectively.
The chambers are required to tolerate the prompt burst and return an
operational state soon after.

Space charge effects are known to limit the performance of wire chambers
in high-rate environments.
When a charged particle passes through a wire chamber, the chamber gas is
ionized, producing initial ion-electron pairs.
As electrons drift and approach anode wires, many ion-electron pairs
are created by avalanche multiplication.
While the electrons are collected quickly by the anode wires, the ions remain
in the chamber for more than ten~$\mu$s until they are collected by
potential wires or cathode electrodes.
When the number of initial ions is large, because of high-intensity incoming
charged particles,
many secondary ions are created by the avalanche,
which
distorts the electric field in the chamber and
suppresses avalanche
processes for delayed particles \cite{HIMAC}.

We developed a special MWPC with a new dynamic control technique
for avalanche multiplication, which reduces space charge effects during
the burst pulses and makes the detector operational soon after them.
We now describe the detector scheme and its verification.

\section{Potential wire high voltage switching}
\label{sec:HVswitching}
\subsection{Conceptual design study}
Space charge effects can be dealt with by sweeping ions out more quickly,
using a gas with faster ion mobility,
or placing the cathode or potential wires nearer to the anode wires.
Another method for handling space charge effects is to suppress ion production during the period of the burst
pulses.
In this paper, we describe the latter method for dynamically controlling gas gains.

\begin{figure}[hbpt]
   \centering
   \includegraphics[width=25mm]{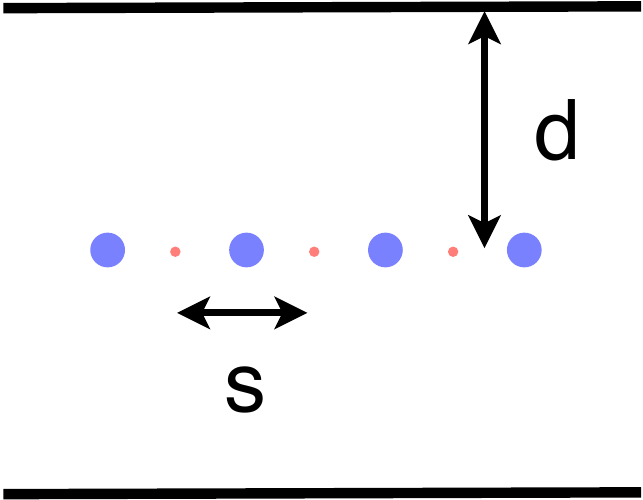}
   \caption{Schematic of the MWPC. Small red dots represent anode wires,
   large blue circles represent
   potential wires, and black horizontal lines represent cathode planes.}
   \label{fig:MWPCStructure}
\end{figure}
\begin{figure}[hbpt]
   \centering
   \includegraphics[width=120mm]{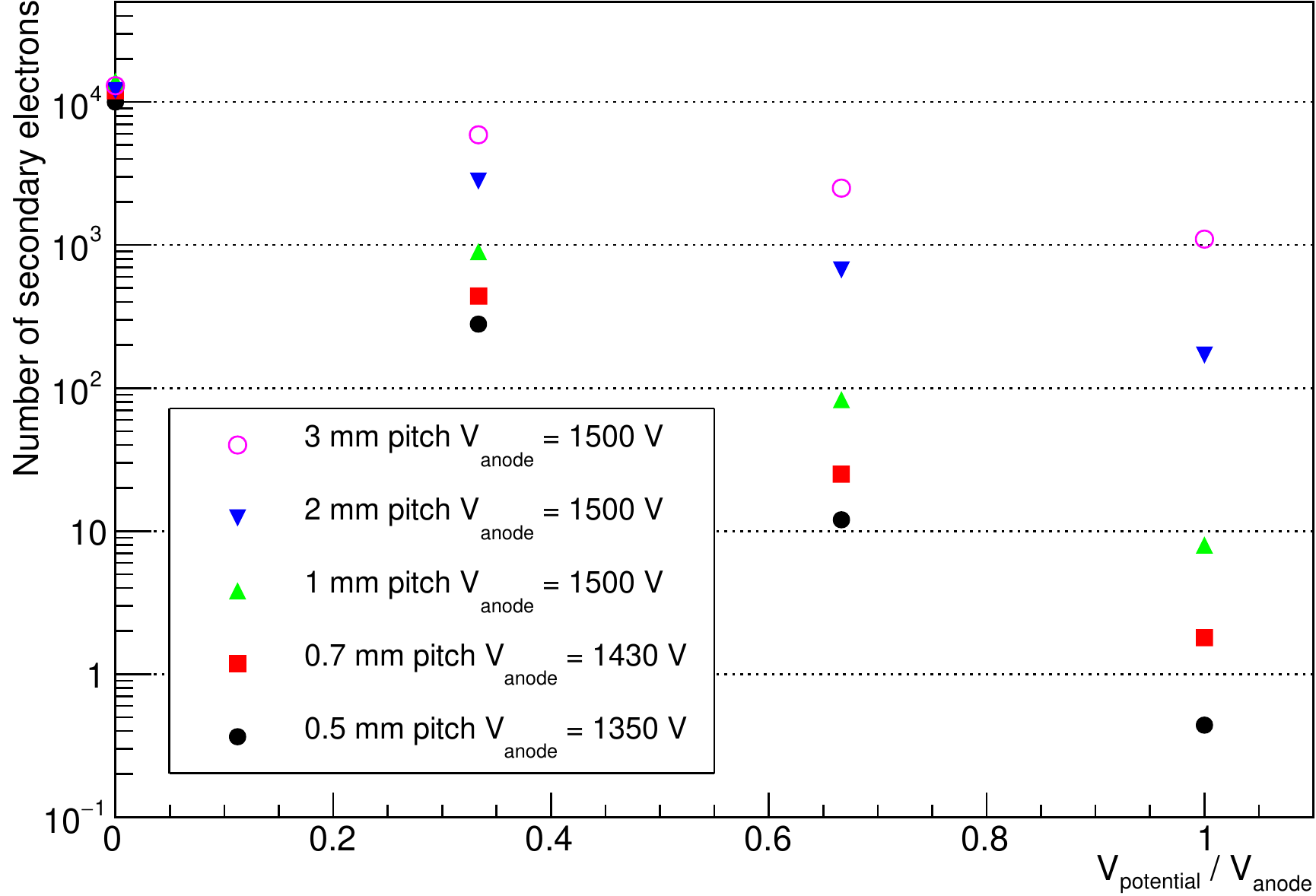}
   \caption{Dependence between the number of secondary electrons produced by
avalanche multiplication and the applied voltage on the potential wires
calculated by GARFIELD++.
The pitch is the distance between an anode wire and a potential wire.
The gap between the wire and the cathode planes is 3~mm.}
   \label{fig:SwitchingGain}
\end{figure}

We consider a wire chamber composed of anode and potential wires placed
alternately in a plane between cathodes, as illustrated in Figure~\ref{fig:MWPCStructure}.
When the wire pitch $s/2$ is sufficiently small compared to the gap between
the wire and cathode planes $d$,
the gas gain is mainly determined by the voltage difference between the anode
and potential wires.
The number of secondary electrons created in avalanche multiplication is
calculated with GARFIELD++~\cite{Garfield++}
and the mean values of 1000 events are shown in Figure~\ref{fig:SwitchingGain}.
In these calculations, the diameters of the anode and potential wires are set
to 15 and 50~$\mu$m, respectively;
$d$ is 3~mm;
and the gas mixture is $\mbox{Ar}/\mbox{C}_2\mbox{H}_6=50/50$.
The applied high voltage (HV) on the anode wires (V$_{\mathrm{anode}}$) is chosen so that the gas gain
approximately equals $1 \times 10^4$ when
the voltage on the potential wires (V$_{\mathrm{potential}}$) is 0 V.
The number of secondary electrons is suppressed when there is
no voltage difference between the anode and the potential wires
and $s \ll d$.

\begin{figure}[h]
	\begin{center}
	\subfigure[
	]{
		\includegraphics[width=0.47\textwidth,clip]{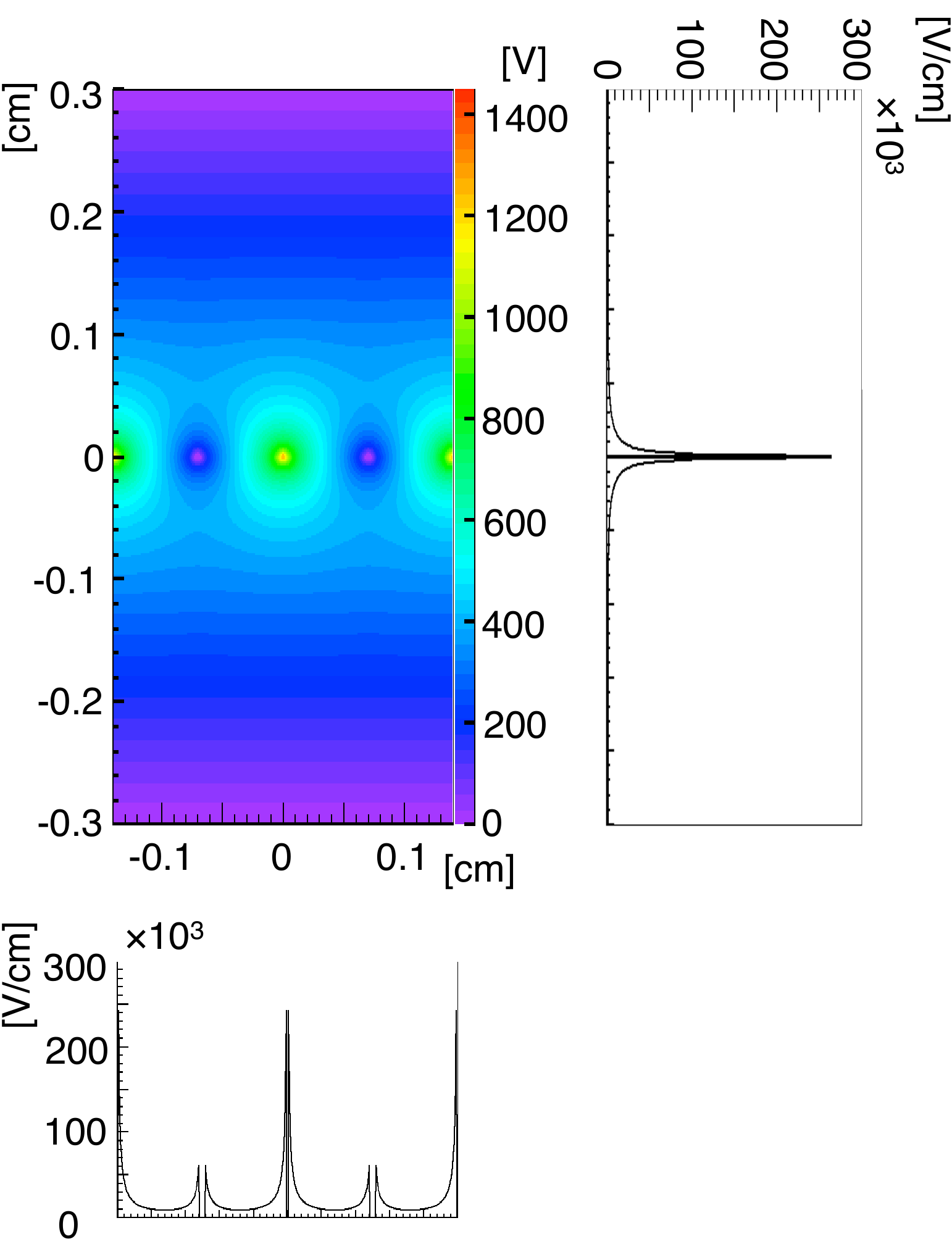}
		\label{contour0}
	}
	\subfigure[
	]{
		\includegraphics[width=0.47\textwidth,clip]{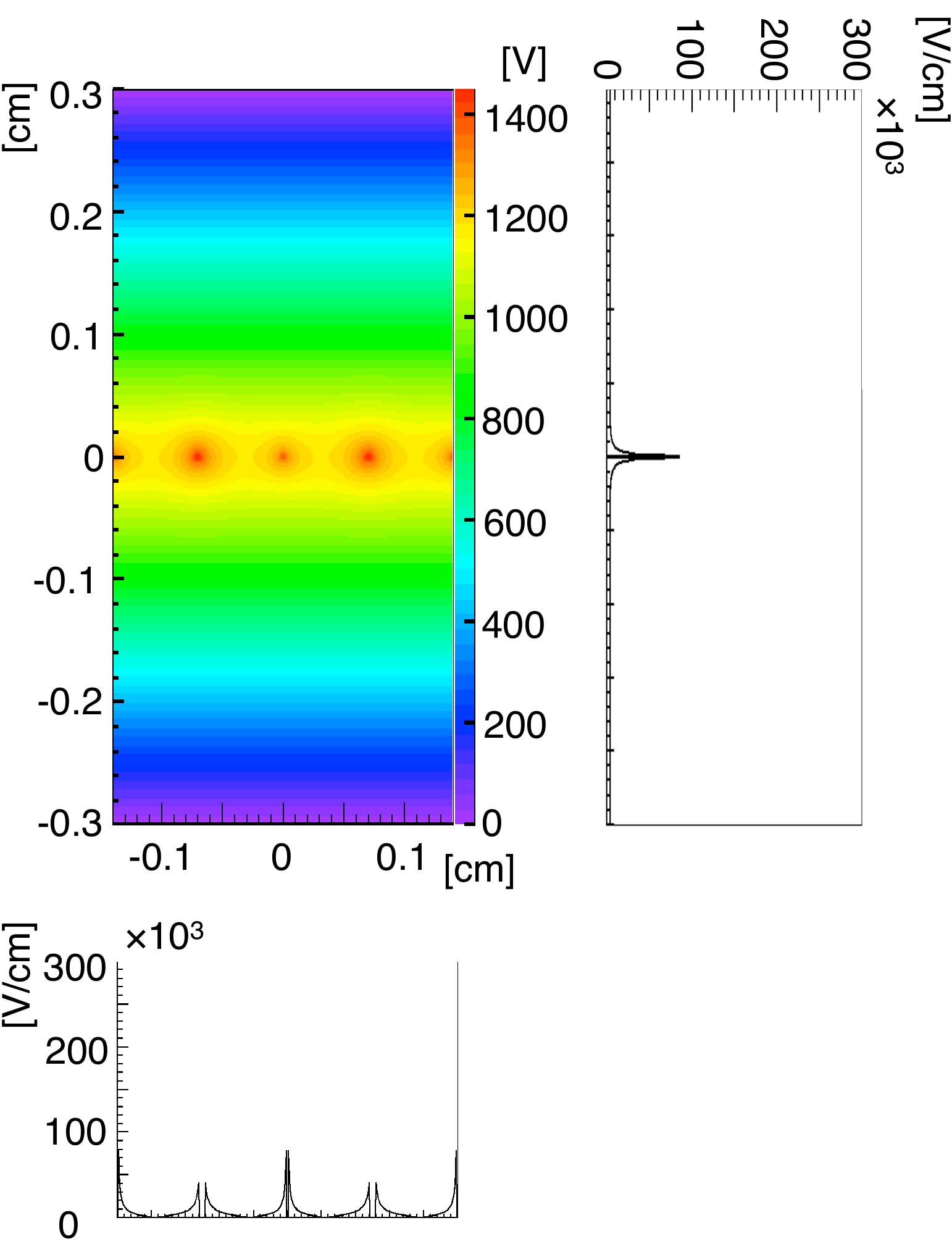}
		\label{contour1000}
	}
	\end{center}
	\caption{
    Contour plots in the upper left of (a) and (b) show the electric potential V of the MWPC
    calculated by GARFIELD++ for the cases in which (a) V$_{\mathrm{anode}}$=1430~V and V$_{\mathrm{potential}}$=0~V,
	and (b) V$_{\mathrm{anode}}$=V$_{\mathrm{potential}}$=1430~V.
    The right and bottom plots around each contour plot show the electric field
	$|\boldsymbol{E}|$ along $x=0$ and $y=0$, respectively.
	}
	\label{fig:fieldcontour}
\end{figure}

The electric potential and electric field calculated for the case in which
$s/2 = 0.7$~mm and $d = 3$~mm are shown in Figure~\ref{fig:fieldcontour}.
The applied voltage is 1430~V on the anode wires and 0~V on the
potential wires
in Figure~\ref{contour0} and 1430~V in Figure~\ref{contour1000}.
The colored contour plots in the Figures~\ref{contour0} and \ref{contour1000}
show the electric potential;
and graphs in the right and the bottom of them
show the electric field profile along $x$ = 0 and $y$ = 0, respectively.
One anode wire is chosen to be located at $(x, y) = (0, 0)$.
The electric potential decreases rapidly as the distance from an anode wire increases when
V$_{\mathrm{potential}}=0$~V.
However, when HV is applied to the potential wires,
the electric potential changes slowly
and the electric field becomes small,
which causes the acceleration that an electron receives within the mean free path near the
wire small enough so that the ionization of gas molecules is suppressed.

\begin{figure}[hbpt]
   \centering
   \includegraphics[width=70mm]{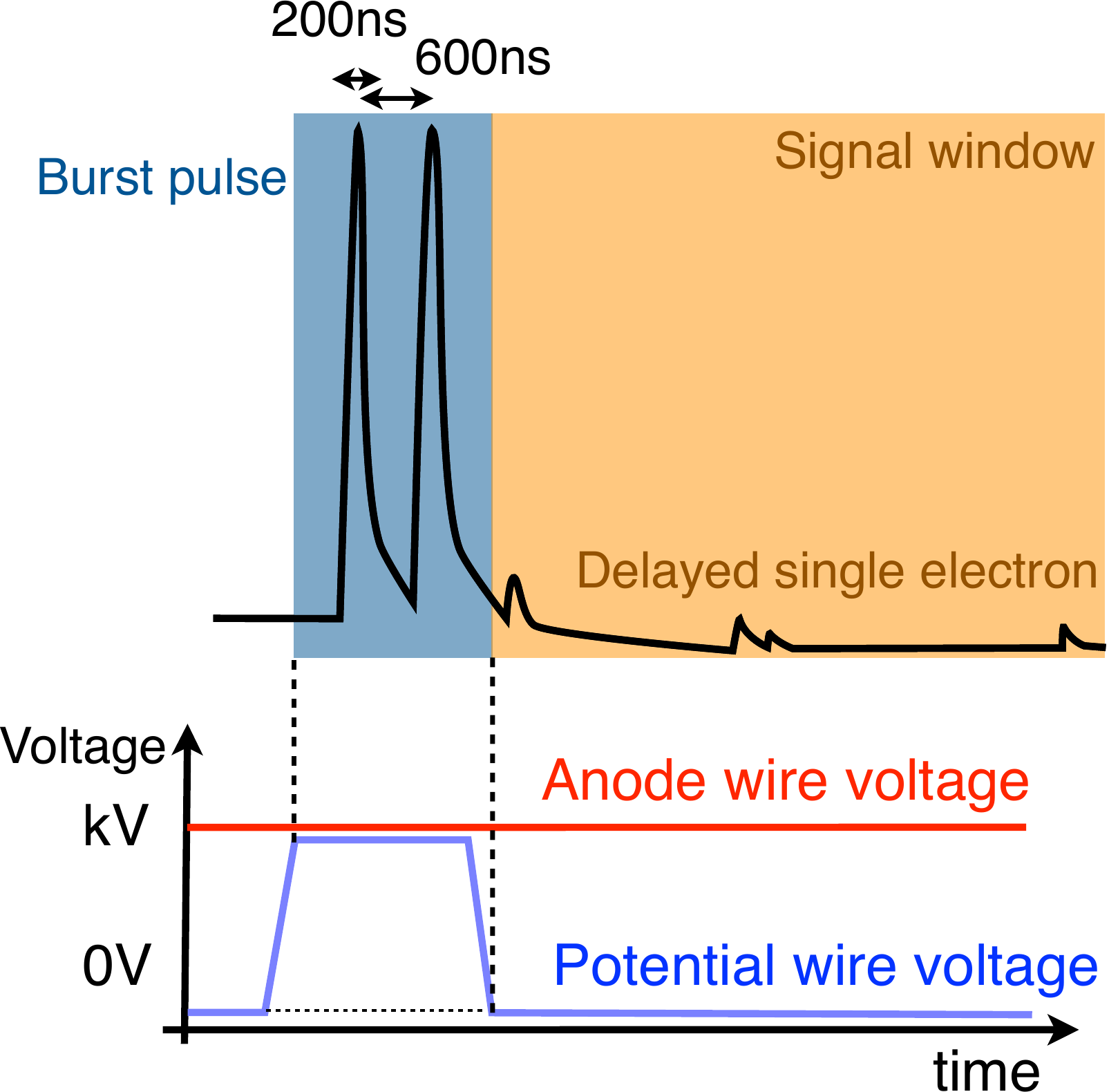}
   \caption{Timing schematic of an expected MWPC output waveform
and HV switching.}
   \label{fig:HVSwitchingScheme}
\end{figure}

Figure~\ref{fig:HVSwitchingScheme} illustrates a scheme to dynamically control gas gain,
which sweeps initial electrons out without gas multiplication.
One might think that rapid voltage changes on the wires would break the
balance between electrostatic force and wire tension,
resulting in significant operational instability.
However, the impulse is very small and stability is maintained if the voltage
switches only for a short period of time---a few microseconds---and then returns to its original state.
Switching the voltage on the potential wires has two advantages compared to
doing so on the anode wires.
One is that initial electrons are swept out quickly.
It is possible to stop avalanche multiplication by
decreasing the voltage on the anode wires to, for example, 0~V,
but the initial electrons remain inside the chamber because there is
no positive electrode to absorb them and
they will invoke avalanche multiplication when the voltage switches back
to its operational value.
Another advantage is related to noise reduction.
We put low-pass filters between the wires and high voltage (HV) power
supply to cut noise from the power supply,
which conflicts with fast voltage switching on the anode wires.
Conversely, the potential wires are grounded during the
signal amplification period
and no noise influence is expected by directly connecting the voltage
supply and potential wires to a common ground.

\subsection{HV switching circuit}
\begin{figure}[hbpt]
   \centering
   \includegraphics[width=65mm]{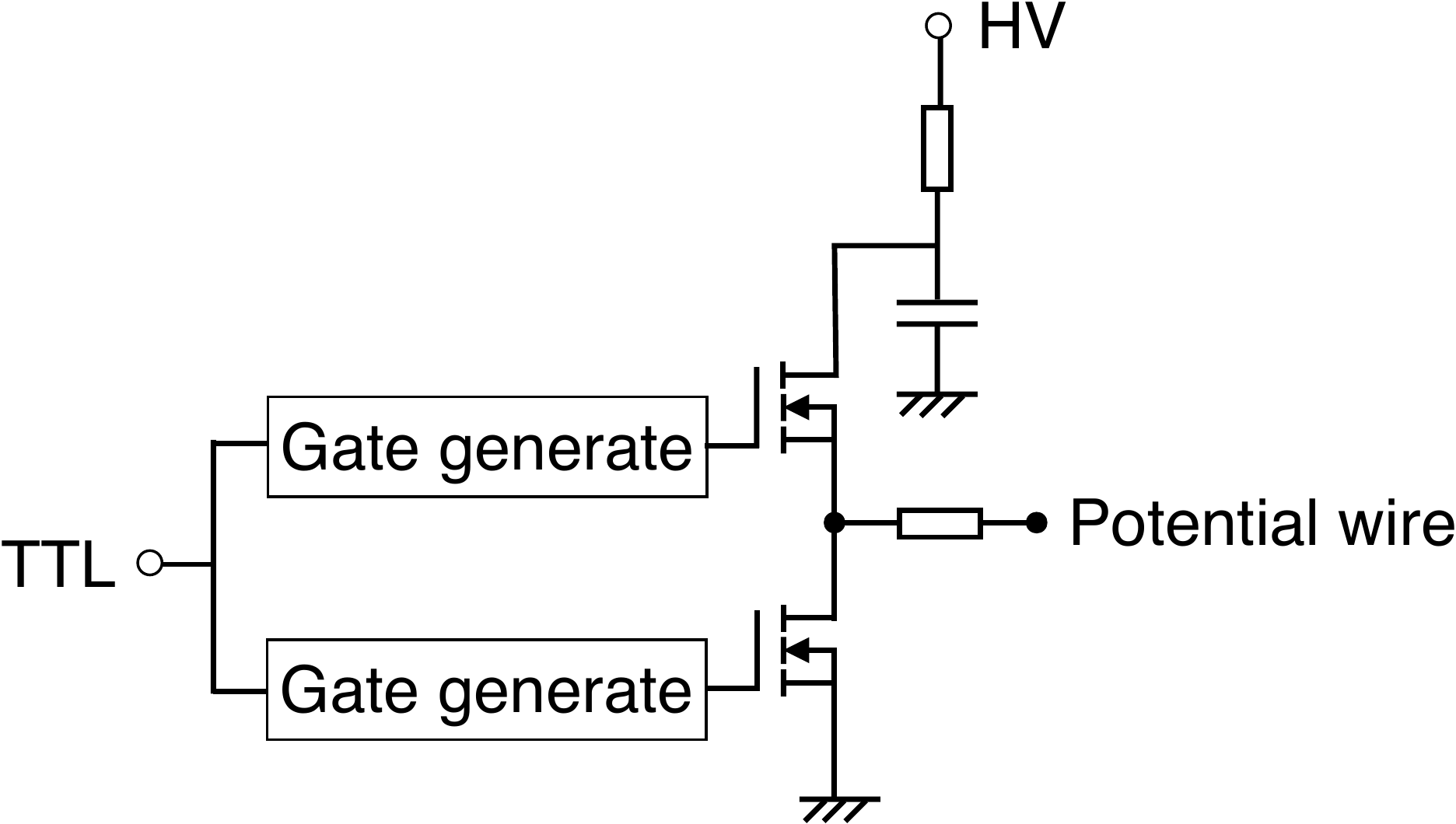}
   \caption{Schematic circuit diagram of HV switching.}
   \label{fig:HVswitchingCircuit}
\end{figure}

The voltage for the potential wires is driven by switching them between the ground
and HV.
We utilized an $n$-channel power-MOSFET WPH4003-1E with a drain-to-source voltage of 1.7~kV
and a drain current of 3~A.
This element works with rise, fall, and turn-on delay times less
than 100~ns and a turn-off delay time of 200~ns.
A schematic design of the circuit diagram for the pulsed HV supply is shown
in Figure~\ref{fig:HVswitchingCircuit}.
When a TTL-high comes in, the lower MOSFET in the figure
is turned off to disconnect the output from the ground,
while the upper MOSFET is turned on to extract charges from a large
capacitance.
When a TTL-low comes in, the connection to the HV is cut and the
potential wires are connected to the ground.
The capacitance is then recharged.
Gate signals are arranged so that both MOSFETs are not turned on at the
same time.

\section{Chamber design}
A smaller wire pitch is desired to effectively suppress gas multiplication, as we
discussed in Section~\ref{sec:HVswitching}.
A smaller cell size also contributes to higher rate tolerance since the number of
passing particles per cell becomes smaller.
However,
a smaller wire pitch results in larger electrostatic forces between the anode and potential wires, as well as
 higher risks of electrical discharges.
We discuss wire sag due to electrostatic forces in Subsection~\ref{subsec:sag} and discharge voltage in Subsection~\ref{subsec:discharge}.

\subsection{Wire sag due to electrostatic force}
\label{subsec:sag}
\begin{figure}[hbpt]
   \centering
   \includegraphics[width=110mm]{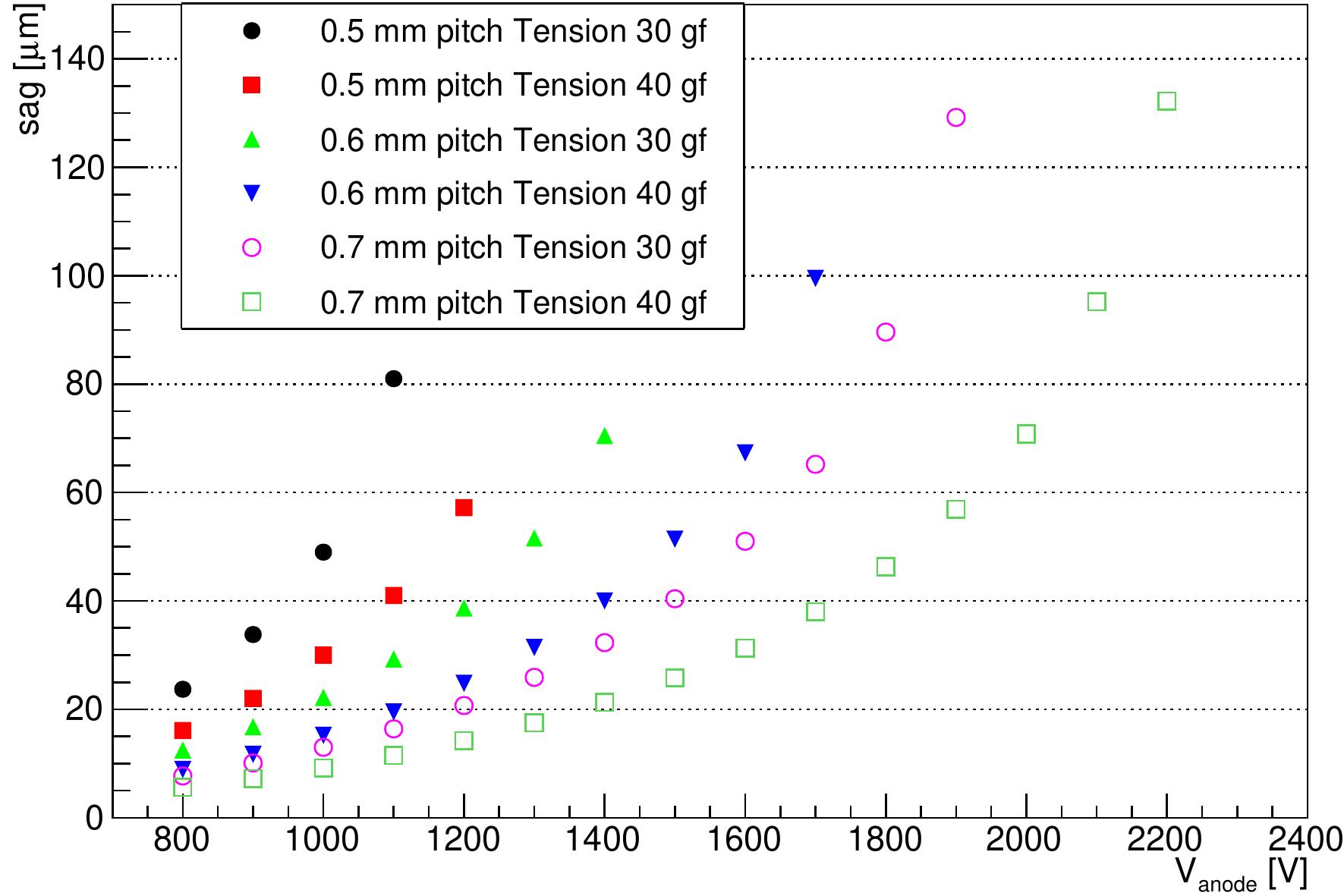}
   \caption{Wire sag due to electrostatic force calculated by GARFIELD.}
   \label{fig:sag}
\end{figure}

When an anode wire is displaced from the center of adjacent potential wires, the balance of electrostatic forces is lost
and the wire should move to new positions in which the electrostatic force and the restoring force (caused by wire tension) balance with each other.
If the applied voltage is too high, the electrostatic force overcomes the restoring force and the anode wire eventually touches an adjacent potential wire.
The expected amount of sag due to electrostatic force is calculated by GARFIELD \cite{Garfield} with an initial displacement of 50~$\mu$m, as shown in Figure~\ref{fig:sag}.
Since the tensile strength at yield point of rhenium-tungsten wire with a diameter of 15~$\mu$m is 51~gf,
the calculations are done with wire tensions of 30 and 40~gf.
The anode wire length is set to 342~mm, assuming the active region to be 300~mm and the
margins to be 21~mm at both ends.
Voltages that should be applied on the anode wires to obtain the nominal gas gain of $1 \times 10^4$ with a gas mixture of
$\mbox{Ar}/\mbox{C}_2\mbox{H}_6 = 50/50$ are calculated to be approximately 1350, 1400, and 1430~V for wire pitches of $s/2=0.5$, $0.6$,
and $0.7$~mm, respectively.
According to Figure \ref{fig:sag}, the sag would be too large for a wire pitch of 0.5~mm to obtain the nominal gain.
The chamber with the wire pitch of 0.7~mm is expected to be operable with wire sag less than 40~$\mu$m even when the wire tension decreases to 30~gf.
The calculation shows that the electrostatic force does not overcome the restoring force
to approximately 2000~V, which is sufficiently larger than the voltage needed to obtain the nominal gain.

\subsection{Discharge voltage measurement}
\label{subsec:discharge}
\begin{figure}[hbpt]
   \centering
   \includegraphics[width=110mm]{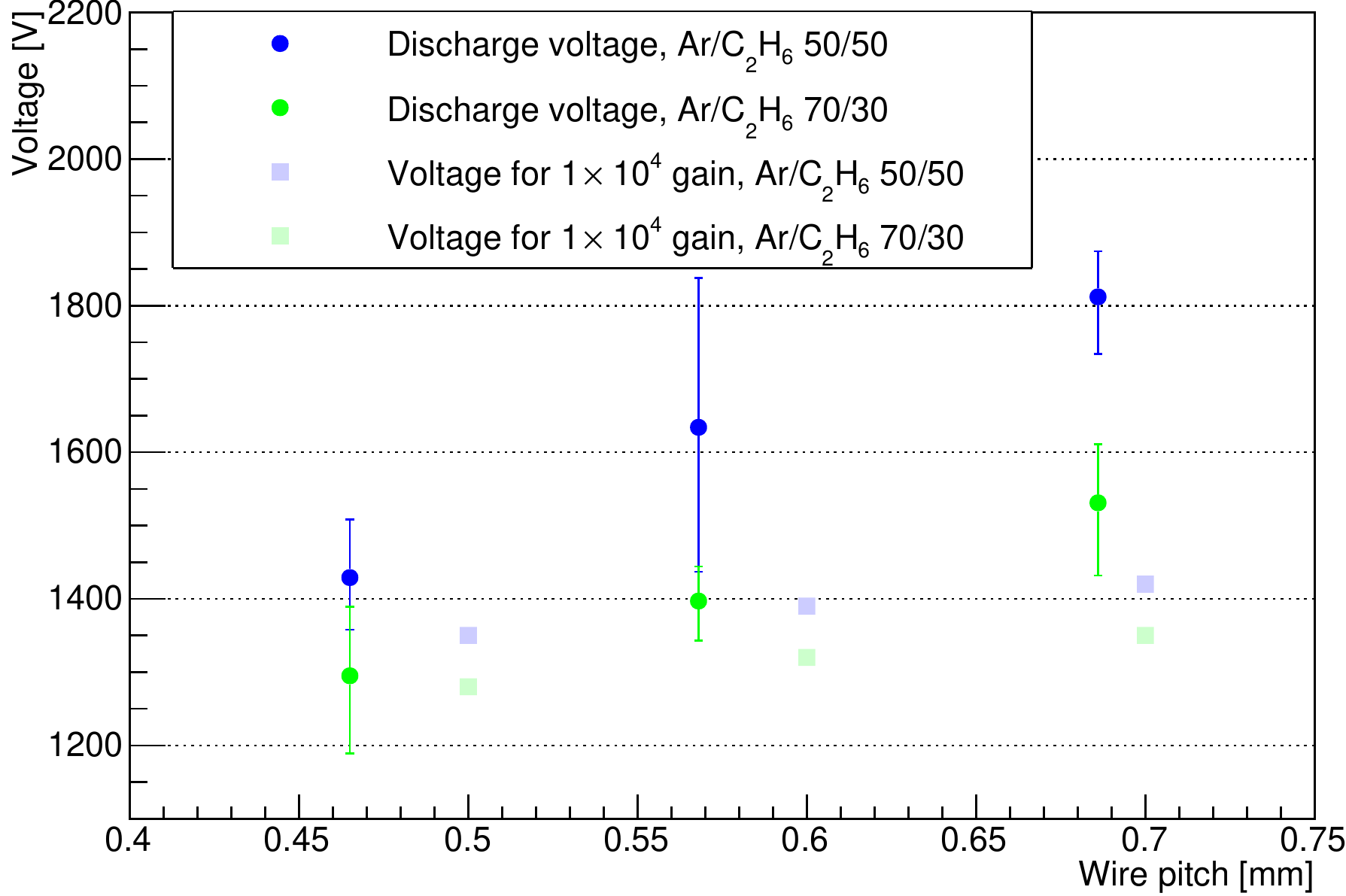}
   \caption{Measured discharge voltage (circles) and required voltage for a gain of $1 \times 10^4$ calculated by GARFIELD++ (squares).}
   \label{fig:discharge}
\end{figure}

\begin{figure}[h]
	\begin{center}
	\subfigure[
	]{
		\includegraphics[width=0.23\textwidth,clip]{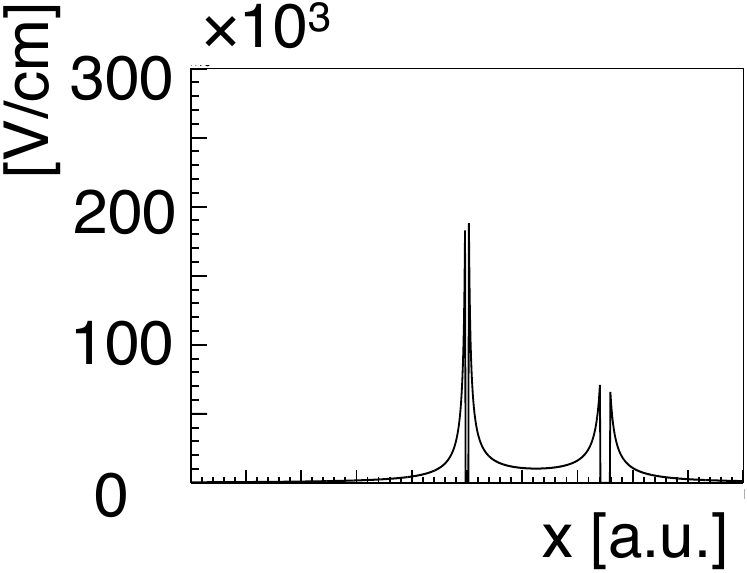}
		\label{EFieldWCath}
	}
	\subfigure[
	]{
		\includegraphics[width=0.23\textwidth,clip]{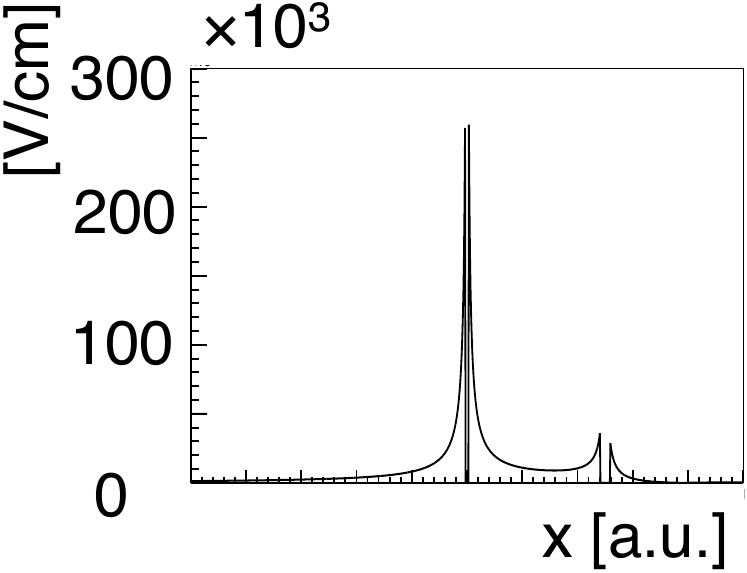}
		\label{EFiedWoCath}
	}
	\end{center}
	\caption{
	Electric field $|\boldsymbol{E}|$ calculated by GARFIELD++ for the two wire configuration with different conditions.
	Two wires with respective diameters of 15 and 50~$\mu$m and respective voltages of 1430 and 0~V are
	placed with a distance of 0.7~mm in both plots.
	(a) shows one anode and one potential wire configuration.
    (b) shows one anode, one potential wire configuration, and 0~V electrodes placed
	3~mm from both wires.
	}
	\label{fig:EField_TwoWire}
\end{figure}

We measured discharge voltages for several different wire pitch and gas mixture setup conditions.
Two wires with diameters of 15 and 50~$\mu$m were put in parallel in a region approximately 1~cm
long on an FR4 frame.
Supporting structures for the wires were designed to prevent creeping discharges on the frame.
For each setup, measurements were taken several times.
The circles in Figure~\ref{fig:discharge} show the mean voltage measurements and
the error bars show the highest and lowest measured discharge voltages.
The square represents the voltage required to obtain a gain of $1 \times 10^4$, as calculated
by GARFIELD++.
There is a discharge risk with a setup of $s/2 = 0.5$~mm.
We can obtain the same gain with smaller voltages by increasing the Ar fraction in the gas,
although the decrease in discharge voltages is larger than that of operation voltages, resulting in smaller safety margins.
However, with larger $\mbox{C}_2\mbox{H}_6$ fractions, higher voltages must be applied,
and it becomes difficult to find suitable MOSFETs for such high operation voltages.
Higher voltages are also undesirable for stability, owing to wire sag.
The electric field in this measurement is not identical to that of the actual chamber.
Figure~\ref{fig:EField_TwoWire} shows the electric field calculated for the two-wire configuration with and without the
cathode plane. The FR4 structure is not included in this calculation. Although the actual electric field in the chamber
would differ from these models depending on the effective location of the electric ground level, the two-wire
configuration models reality sufficiently well, as we infer from Figure~\ref{contour0} and \ref{fig:EField_TwoWire},
and measurements with this configuration are still useful for obtaining a better understanding
of suitable chamber geometry.

\subsection{Chamber geometry}
A smaller cell size is desirable for a high tolerance rate.
When the gap between the wires and cathode electrodes is smaller than
3~mm, we make the wire pitch less than 0.7~mm, as is inferred from
Figure~\ref{fig:SwitchingGain}. However, as the wire pitch decreases, the risk of wire sag due
to electrostatic force and the risk of discharge increase.
Considering the calculations and measurements discussed in the previous subsections,
we decided to adopt a wire-cathode gap of 3~mm and
an anode-potential wire pitch of 0.7~mm.
\section{Chamber tests}

\subsection{Prototype chamber}
\begin{figure}[hbpt]
   \centering
   \includegraphics[width=80mm]{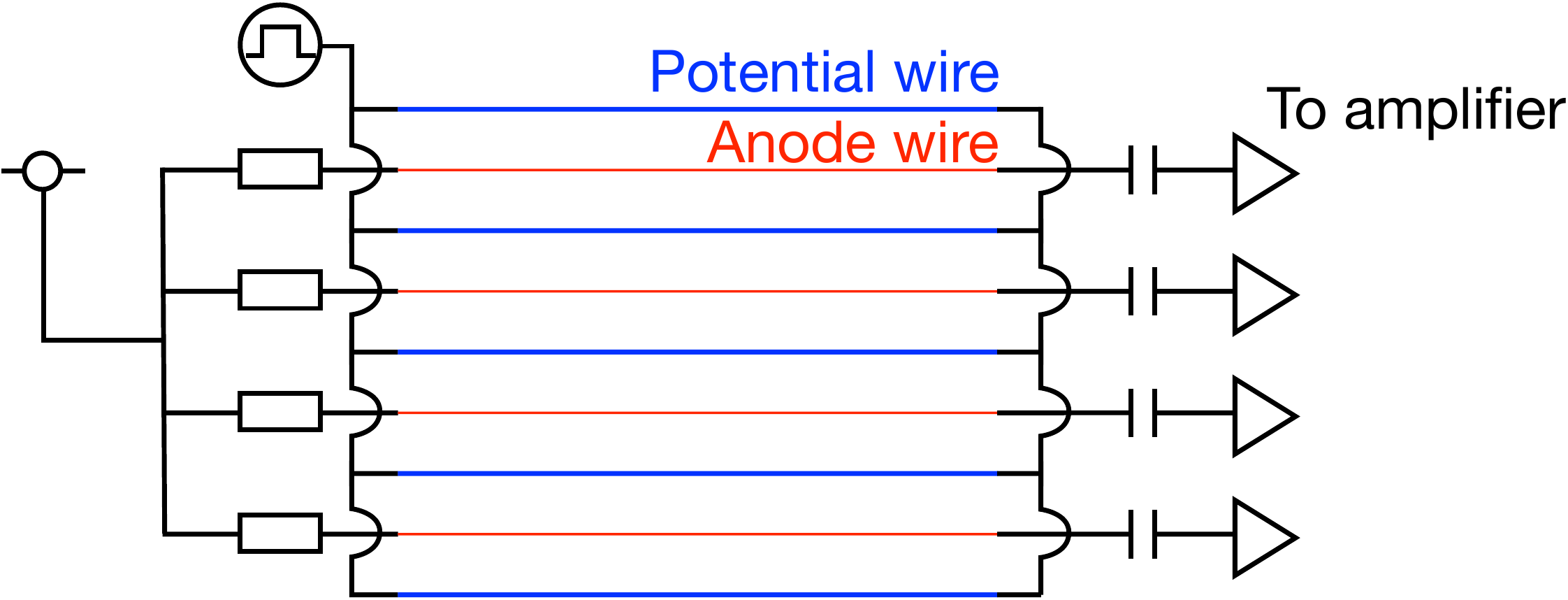}
   \caption{Schematic circuit diagram of the prototype chamber.}
   \label{fig:MWPCCircuit}
\end{figure}

We constructed a prototype chamber containing
 anode and potential wires placed alternately in a plane.
The pitch between the anode and potential wires is 0.7~mm.
The gap between the wire plane and the cathode plane is 3~mm.
The diameters of the anode and the potential wires are 15 and 50~$\mu$m, respectively.
Tension on the wires is approximately 40 and 80~gf for the
anode and potential wires, respectively.
The wire length in the active region is 300~mm.
One side of the cathode is made of a flat aluminum foil, while the other
side is made of an aluminized film with readout strip patterns on it.
We used a gas mixture of $\mbox{Ar}/\mbox{C}_2\mbox{H}_6 = 50/50$.
A schematic circuit diagram of the prototype chamber is shown in Figure~\ref{fig:MWPCCircuit}.
The resistance and capacitance connected to the anode wire were determined to be 2~M$\Omega$
and 2~nF, respectively. The capacitance was later altered to 10~nF, which is described in
Subsection~\ref{subsec:Kurri}.

\subsection{Verification of dynamic gain control}
\begin{figure}[hbpt]
   \centering
   \includegraphics[width=120mm]{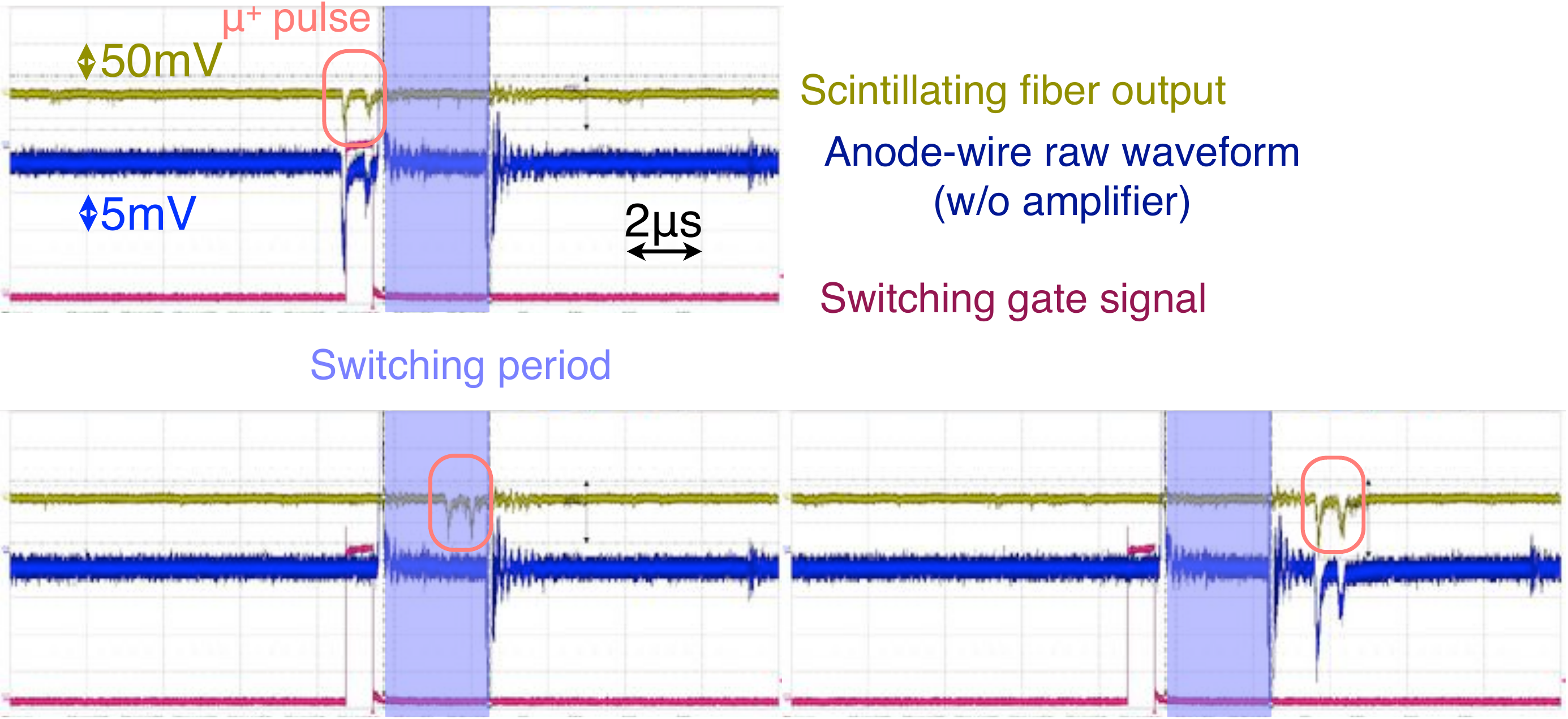}
   \caption{Waveforms of the chamber, a reference detector, and a gate signal for the pulsed HV
   power supply. The chamber output is
   read out without amplifiers. We apply the HV on the potential wires during the period illustrated as
   the hatched region, while the voltage is set to 0 in other periods.}
   \label{fig:MLFBeamtest}
\end{figure}

\begin{figure}[hbpt]
   \centering
   \includegraphics[width=110mm]{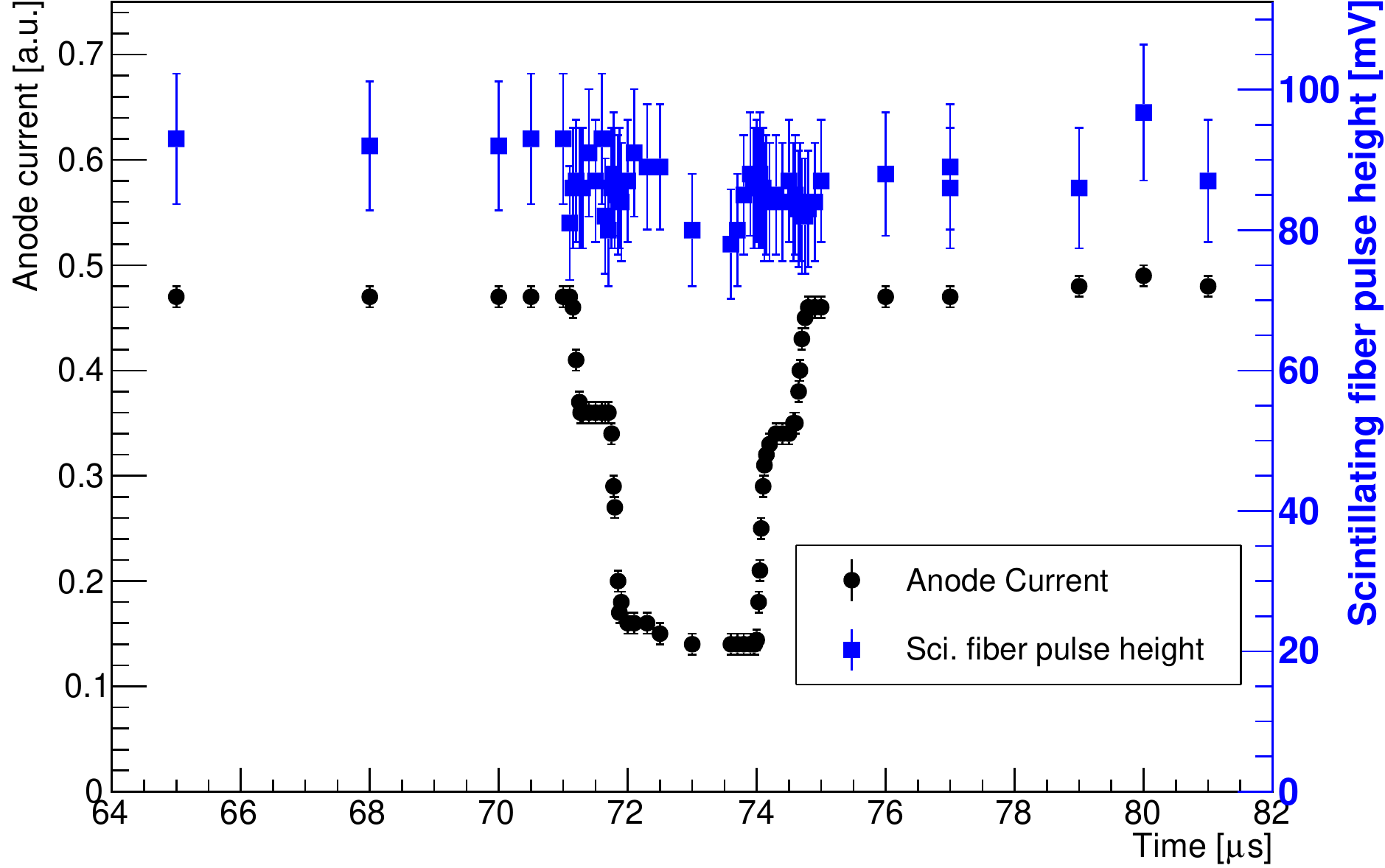}
   \caption{Current readout of the HV power supply for the anode wires
as a function of HV switching timing for the potential wires (solid circles).
The $x$-axis is the delay of the TTL-gate signal for the pulsed HV supply
with respect to a timing signal from the accelerator.
Pulse height of the scintillating fiber is shown as a reference for the
beam intensity (solid squares) with values on a right vertical axis.
The upper, lower left, and lower right
images in Figure~\ref{fig:MLFBeamtest} correspond to approximately
75, 72, and 70 $\mu$s in this figure, respectively.}
   \label{fig:SwitchingCurrent}
\end{figure}

We verified the dynamic gain control by switching the HV
on the potential wires
using double-pulsed muon beams available at the D2 beamline of the MUSE.
Decay muon beams penetrated the chamber
and a scintillating fiber that was read out by a multipixel photon counter.
Switching the HV on the potential wires was controlled by changing
the timing and pulse width of the TTL-gate signal sent to the pulsed HV
power supply.
Figure~\ref{fig:MLFBeamtest} shows a raw output waveform from an anode wire.
Figure~\ref{fig:SwitchingCurrent} shows the current consumption of the HV supply for the anode wires.
The $x$-axis represents the delay of the gate signal sent to
the pulsed HV supply with respect to a timing signal from the accelerator.
Different $x$ values thus correspond to different relative timings between HV switching
and the main beam double pulses.
The pulse height of signals from the scintillating fiber detector is
also shown in the figure as a reference for the muon beam intensity.
The beam intensity fluctuated slightly and there is a small correlation between the
anode current and the pulse height of the scintillating
fiber. The step-like
structure of the anode current is not related to the beam intensity.
When the delay time was less than 71~$\mu$s or greater than 75~$\mu$s, both muon
double pulses were amplified, so the current consumption was high.
When one of the two pulses was within the switching period,
a step in the current consumption plot occurred.
When both pulses were within the switching period, signal
multiplication of the muon pulses was
suppressed and current consumption was low.
Gain suppression by the HV switching was confirmed and
behavior after the switch was as stable as that before it.

\subsection{Delayed electron detection}
\label{subsec:Kurri}
\begin{figure}[hbpt]
   \centering
   \includegraphics[width=120mm]{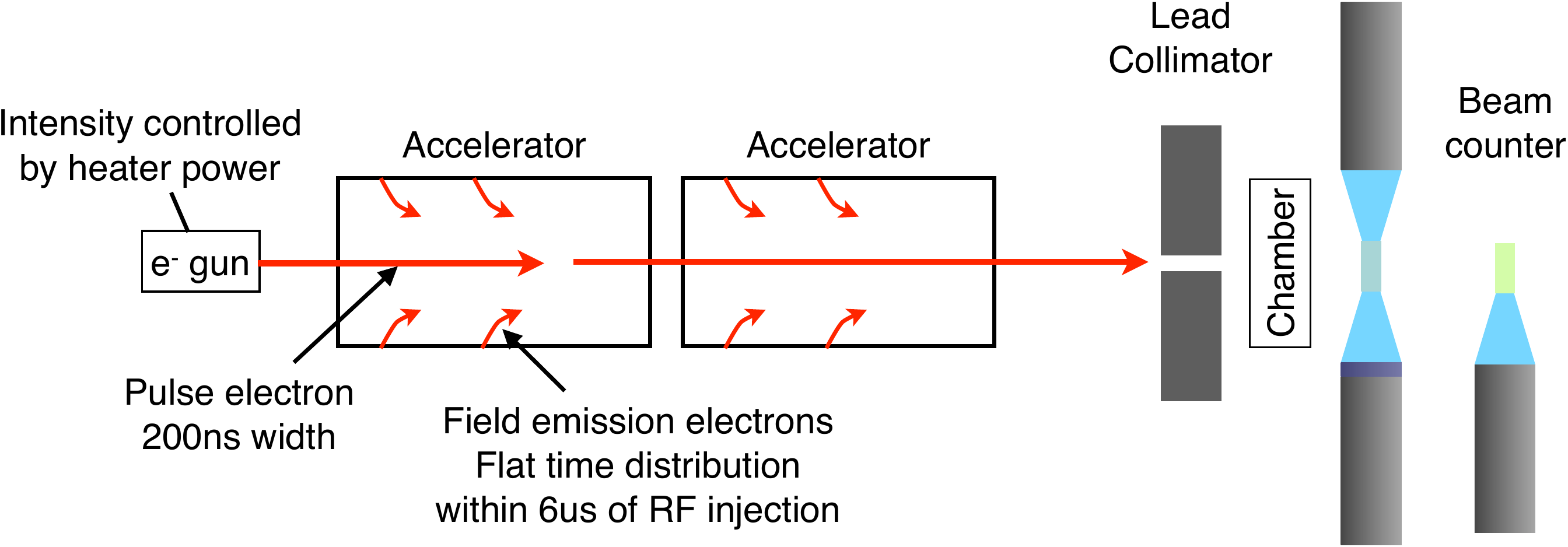}
   \caption{Beam test experiment setup.}
   \label{fig:KURRISetup}
\end{figure}

This test was performed using the Linear Accelerator at
the Kyoto University Research Reactor Institute (KURRI-LINAC).
The experimental setup is shown in Figure~\ref{fig:KURRISetup}.
KURRI-LINAC offers pulsed electron beams whose intensity is controllable
by changing the heater power of its electron gun.
The width, momentum, and repetition of beams were tuned to 200~ns,
20~MeV/$c$, and 25~Hz, respectively.
The distribution of beam intensity was measured to be flat in a collimated
18~mm $\times$ 20~mm region.
In addition to the electrons from the beam gun,
field emission electrons emerging from the surface of the radio frequency
(RF) cavity are accelerated during a period of 6~$\mu$s in which the RF
cavity is filled.
Field emission electrons were utilized to emulate delayed electron signals.
The intensity of the main pulse was measured by beam counters placed
behind the chamber.
A plastic scintillator read out by a photomultiplier tube (PMT) was used
for low intensity beams,
while for higher intensity beams,
acrylic boards read out by PMTs were used to see Cherenkov lights.
One of these had
a neutral density filter with an attenuation of $10^{-3}$ attached to
the PMT, to measure the highest beam intensity.

We reduced the chamber gain during the period in which the main pulse hit
the chamber by switching the HV on for the potential wires;
we switched it off soon after the main pulse passed through the chamber
to detect field emission electrons.
HV switching induces large currents in readout signals from the chamber.
We developed an amplifier with high tolerance against large input currents,
whose impedance is approximately 100~$\Omega$.
\begin{figure}[hbpt]
   \centering
   \includegraphics[width=120mm]{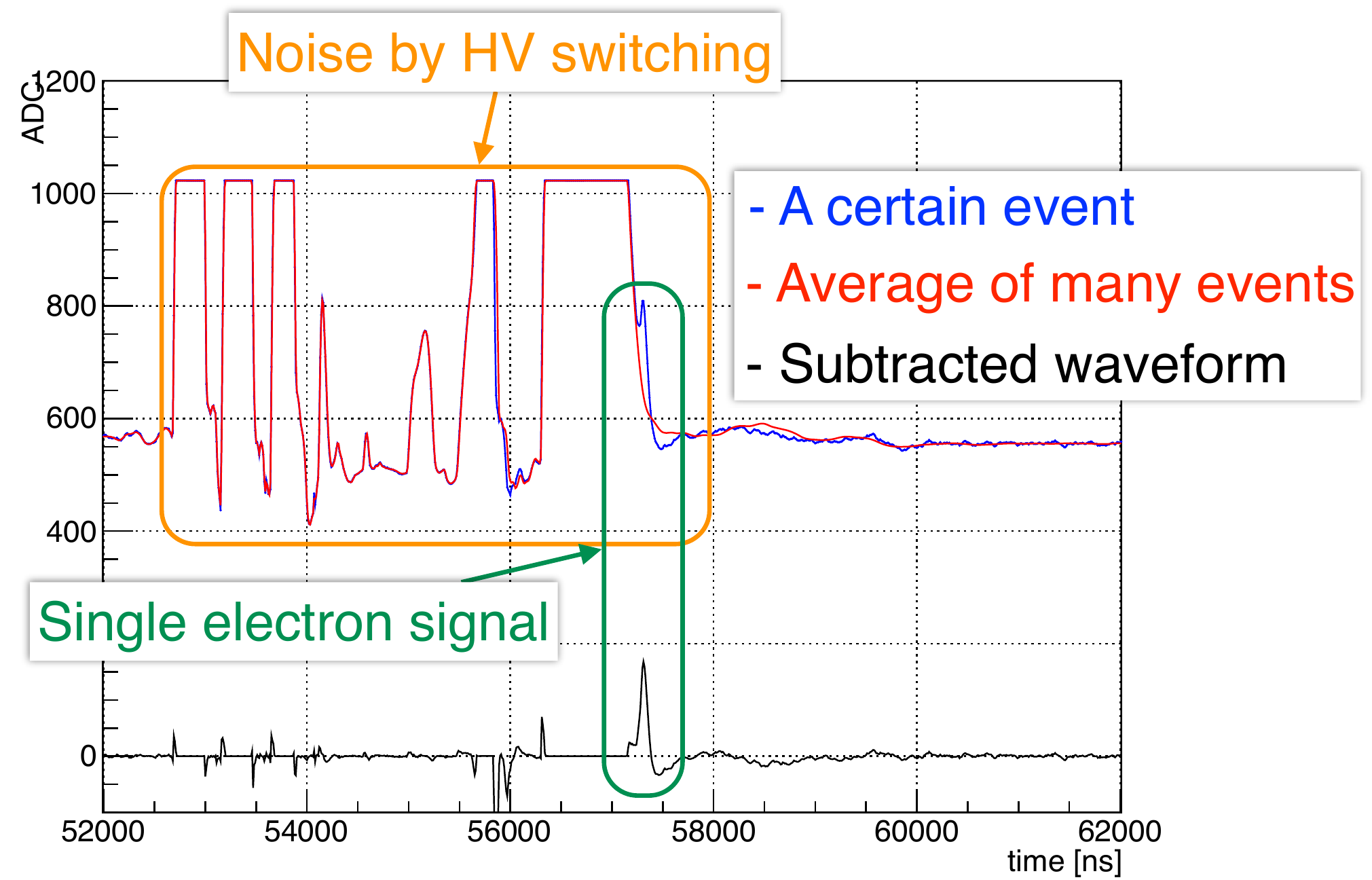}
   \caption{MWPC output waveforms with a burst pulse around 56000 ns
and the HV switching operation for the potential wires.
The noise from HV switching is always the same, so subtracting the
noise gives a clear waveform for the signal electron.
Beam intensity fluctuation appears at around 56000 ns.
Space charge effects do not degrade detector performance and a delayed
electron signal is observed.}
   \label{fig:KURRIResult}
\end{figure}

Resonances were observed on the anode wire and cathode strip outputs
after the amplifier system,
which disappeared when the readout capacitance at each anode wire was
increased from 2 nF to 10 nF.
Since the damping factor for RLC circuits is given by
\begin{displaymath}
\zeta = \frac{R}{2}\sqrt{\frac{C}{L}},
\end{displaymath}
greater damping by the increased capacitance eliminated the resonances.

A typical MWPC waveform outputs for a beam intensity of
70 GHz/mm$^2$ and a width of 200 ns
is shown in Figure~\ref{fig:KURRIResult}.
The noise, with saturated outputs
induced by HV switching,
always shows the same waveform, which enabled us to obtain clean signals
by subtracting an averaged waveform.
An electron signal can be clearly observed
after the burst pulse, as shown in the figure.

\section{Summary}
We are searching for $\mu$-$e$ conversions with a sensitivity of $\mathcal{O}(10^{-14})$
with a new experiment, DeeMe, at the MLF at J-PARC.
The detector must tolerate a double-pulsed burst of electrons with an
instantaneous hit rate of 70 GHz/mm$^2$ in a time width of 200 ns
and is operational for detecting delayed electrons.
We developed an MWPC with a special configuration and a new
technique to suppress space charge effects by sweeping electrons out and
dynamically controlling gas multiplication.
This technique was confirmed to successfully function in the conditions
of the DeeMe experiment.
\section{Acknowledgements}
This work was supported by JSPS KAKENHI grant number JP24224006.
Part of this work was performed using facilities at J-PARC
and Kyoto University Research Reactor Institute.
We thank the staff of the beam facilities for their support during test experiments,
especially N. Kawamura, Y. Miyake of J-PARC, N. Abe and T. Takahashi of KURRI-LINAC.
We appreciate H. Okuma, E. Hirao, K. Konno and S. Maki of REPIC for great works on designing and constructing
a novel MWPC.

\end{document}